\pdfoutput=1
\documentclass[a4paper,10pt]{article}

\usepackage{graphicx}
\usepackage{psfrag}
\usepackage{subfigure}
\usepackage[colorlinks=true
,urlcolor=blue
,anchorcolor=blue
,citecolor=blue
,filecolor=blue
,linkcolor=black
,menucolor=blue
,pagecolor=blue
,linktocpage=true
]{hyperref}

\usepackage[super,nonamebreak]{natbib}
\usepackage{amsmath}
\usepackage{amssymb}
\usepackage{color}
\usepackage{epic,eepic,rotating}

\title{\textbf {\begin{LARGE}Solving Hodgkin-Huxley equations using the compact difference scheme - somadendrite\end{LARGE}}}

\author{\begin{normalsize}%
  Asha Gopinathan\thanks{PI,Dendritic Simulator project, 
     SCTIMST,Trivandrum - 695011, Kerala, India, 
     Mob: +919633568106, E-mail: dendron.15@gmail.com} 
     \&  Joseph Mathew\thanks{Department of Aerospace Engineering, 
     Indian Institute of Science, Bangalore 560012, India} 
\end{normalsize}} 

\date{ }

\begin{document}

\maketitle       
\title{}
\author{}
\newpage
\section*{\begin{large}\textbf{Summary}\end{large}}
Dendrites have voltage-gated ion channels which aid in production of action potentials. Thus dendrites are not just passive conductors of information, but actively act on the incoming input.  Here we assume Hodgkin-Huxley formulations of voltage-gated ion channels on the dendrite. These equations are normally solved by some form of central difference scheme or the spectral methods. We use a compact finite-difference scheme to solve these equations. This scheme gives spectral-like spatial resolution while being easier to solve than spectral methods.  The scheme has shown to be able to reproduce the results from spectral methods. In this paper cylindrical dendrites are described. It may also be noted that the compact difference scheme can be used to solve any other PDE both in biological as well as nonbiological systems. It is increasingly used in studying turbulence in airflow. 

\section*{\begin {normalsize}\textbf{Introduction}\end{normalsize}}
Dendrites are long extensions to neurons which receive both excitatory and inhibitory synaptic inputs$^{1}$. These are received on dendritic spines$^{2,3,4}$, or branches ( mostly excitatory ) and dendritic shaft ( inhibitory). The integration of multiple inputs can result in many types of action potentials - those that move towards the soma from the synapses on the  dendrite and others that move into the dendrite from the soma (back-propagation)$^{5,6,7,8}$. This can be modeled by thinking of the dendrite as a cable with ion channels distributed along it. It is the nonlinearity caused by these ion channels$^{9}$(both type and distribution ) that result in a range of action potential shapes with different firing frequencies. \\
\parindent = 1in The change in voltage with respect to time and space can be modeled with the cable equation for passive dendrites$^{10,11,12}$. But in the case of active dendrites, the Hodgkin-Huxley equations are used$^{11,12,13}$. Assuming that there are just sodium, potassium and leak channels, these equations take the following form 
\begin{normalsize}
\begin{equation}
 C_{m}\frac{\partial{V}}{\partial{t}} = \gamma_{0}(x)\frac{\partial^{2}V}{\partial{x}^2} + \gamma_{1}(x)\frac{\partial V}{\partial x}-I_{ion} + I_{in}(x,t)
\end{equation}
\begin{equation}
\gamma_{0}(x) = \frac{1}{2R_{i} }\frac{ r(x)}{\sqrt{(1+r'^{2}(x))}}
\end{equation}
\begin{equation}
 \gamma_{1}(x)= \frac{1}{R_{i}}\frac{ r'(x)}{\sqrt{(1+r'^{2}(x))}}
\end{equation}
where $r$ is the radius of the dendrite, $r'(x) = dr/dx$, $C_{m}$ is the constant membrane capacitance,$R_{i}$ is the constant axial resistivity,$I_{ion}= I_{Na}+I_{K}+I_{L}$ and $I_{in}$ is the injected current.\\
\begin{equation}
 I_{Na}= g_{Na}(x)m^{3}h(V-E_{Na}), I_{K}= g_{K}(x)n^{4}(V-E_{K}), I_{L}= g_{L}(V-V_{L})
\end{equation} 
Equations for evaluating $g_{Na}$ and $g_{K}$ are given in reference $14$.\\
The temporal change of the potassium activation particle $n$ and sodium activation particle $m$ and inactivation particle $h$ are given by :
\begin{equation}
 \frac{dn}{dt} = \alpha _{n}(V)(1-n) - \beta _{n}(V)n
\end{equation}
\begin{equation}
 \frac{dm}{dt} = \alpha_{m}(V)(1-m)-\beta_{m}(V)m
\end{equation}
\begin{equation}
 \frac{dh}{dt} = \alpha_{h}(V)(1-h)-\beta_{h}(V)h
\end{equation}
$\alpha_{m}$, $\beta_{m}$, $\alpha_{h}$, $\beta_{h}$, $\alpha_{n}$, $\beta_{n}$ are evaluated from formulae given in reference $14$.\\ 
When the dendrite is cylindrical equation $1$ reduces to :
\begin{equation}
 C_{m}\frac{\partial{V}}{\partial{t}} = \gamma_{0}(x)\frac{\partial^{2}V}{\partial{x}^2} -I_{ion} + I_{in}(x,t)
\end{equation}
Writing in nondimensional terms when current is injected at $x = 0$,:
\begin{equation}
\frac{ C_{m}}{\tau_{m}}\frac{\partial V}{\partial T} = \frac{\gamma_{0}(X)}{\lambda^{2}}\frac{\partial^{2} V}{\partial X^{2}} - I_{ion}
\end{equation} 
Here, $T= t/\tau_{m}$, $X=x/\lambda$, $\tau_{m}= R_{m}C_{m}10^{3}$ msec, $\lambda = (1/2)(d/(\pi f R_{i}C_{m}))^{1/2}$ cm,\\$f = 1/(2 \pi \tau 10^{-3})$ Hz, $d$ is the diameter in cm, $R_{m}$ is the membrane resistance in $\Omega$cm$^{2}$, $C_{m}$ is the membrane capacitance in $Farad/cm^{2}$ and $R_{i}$ is $\Omega$.cm.
 
\end{normalsize}
\section*{\begin{normalsize}\textbf{Spatial discretisation : Using compact finite difference schemes to solve the cable equation}\end{normalsize}}
\parindent = 1in The second derivative $\partial^{2}V/\partial x^{2}$ in equation $(8)$ is approximated using the following formula(ref.$15$, equation $2.2$):
\begin{eqnarray}
\beta V''_{i-2} + \alpha V''_{i-1} + V''_{i} + \alpha V''_{i+1} + \beta V''_{i+2}& = & \frac{c (V_{i+3}-2V_{i}+V_{i-3})}{9h^{2}}\nonumber \\ & &+\frac{b(V_{i+2}-2V_{i}+V_{i-2})}{4h^{2}} \nonumber \\
& & +\frac{ a( V_{i+1}-2V_{i}+V_{i-1})}{h^{2}}, \\& &  (2\leq i\leq N -1 ) \nonumber 
\end{eqnarray} 
where $V''_{i}$ represents the finite difference approximation to the second derivative at node $i$ and $N$ is the maximum number of nodes in any given grid.
 The relations between the coefficients a,b,c and $\alpha$ , $\beta$ are derived by matching the Taylor series coefficients of various orders. 
We take (ref.$15$,equation $2.2.7$). 
\begin{displaymath}
\alpha = \frac{2}{11}, \beta = 0, a = \frac{12}{11}, b =\frac{ 3}{11}, c= 0
\end{displaymath}
to obtain a sixth order formula.The truncation error is $\frac {-23}{55440}h^{6}d^8V/dx^8$ which is sixth order accurate(ref.${15}$,Table $II$) For the boundaries the scheme chosen is (ref.$15$,equation $4.3.4$)
\begin{equation}
V''_{1}+ \alpha V''_{2} = \frac{ aV_{1}+bV_{2}+cV_{3}+dV_{4}+eV_{5}}{h^{2}}
\end{equation} 
A similar equation connects $V''_{N}$ and $ V''_{N-1}$. 
By requiring third - order formal accuracy the coefficients are reduced to (ref.${15}$,equation $4.3.6$).
\begin{eqnarray*} 
a = \frac{(11 \alpha + 35)}{12}, \hspace{2mm} b = \frac{- (5\alpha +26)}{3} ,\hspace{2mm} c = \frac{(\alpha +19)}{2}, \hspace{2mm} d =\frac{(\alpha -14)}{3}\\  e = \frac{(11 - \alpha)}{12}
\end{eqnarray*}
Choosing $\alpha=1/10$ specifies all these coefficiants from classical Pade scheme which is fourth order. The truncation error is reduced to $ ((\alpha -10)/(12))h^{3}d^5V/dx^5$. If $\alpha$ is $10$, truncation error becomes $h^{4}$. Equations $10$ and $11$ applied at interior points results in a matrix problem $\textbf{A}V'' = \textbf{B}$ where A is tridiagonal and $V''$ can be obtained easily.
\section*{\begin{normalsize}\textbf{Time discretisation}\end{normalsize}}
 The values for $V''$ calculated from the compact-difference scheme were used to integrate the result in time using an explicit time stepping scheme - forward Euler. If $ T = n\Delta T$, $V^{n} \equiv V(T)$ and $V^{n+1} \equiv V(T+ \Delta T)$, then :
\begin{equation}
 V^{n+1}= V^{n} + f(V^{n},n\Delta T)\Delta T
\end{equation} 
 Stability requires the choice of the time step to be
\begin{equation}
  \Delta T < \frac{\Delta X^{2}C_{m}\lambda^{2}}{\tau_{m}\gamma_{0}(X=0)}                                                                                                                                                                                                                                                                                                                                                                                                                                                                                                                                                               
  \end{equation} 
$\gamma_{0}(X=0)$ is maximum over the dendrite. $\Delta T$ varies as shown in (Table~\ref{tab:DeltaT}).The numerical integration in time has been done with an explicit scheme. Since spatial derivatives are obtained with a compact scheme, which is an implicit formula that requires the solution of a linear system, implicit time-stepping is not possible. Implicit time-stepping is desirable to overcome the severe restrictions that stability imposes on the time-step of conditionally stable explicit schemes. A work-around is to use a predictor-corrector scheme which uses an explicit step estimate from the predictor step in a corrector step which is also an explicit step.Computations were performed on a Toshiba Satellite Pro laptop using Octave in a Linux(Ubuntu)environment. The data used for  simulations is given in (Table~\ref{tab:parameters}) and caption of (Fig.~\ref{fig:cabunbact1}).   
\section*{\begin{normalsize}\textbf{Initial and boundary conditions}\end{normalsize}}
The initial and boundary conditions are given in reference $14$:\\
\begin{equation}
 V(x,0)= V_{o}(x)
\end{equation} 
Here $V_{o}(x)= -60 $ mV
\begin{equation}
 b_{11}V(0,t) + b_{12}\frac{\partial V(0,t)}{\partial x} = f_{b1}(t)
\end{equation} 
\begin{equation}
 b_{21}V(L,t) + b_{22}\frac{\partial V(L,t)}{\partial x} = f_{b2}(t)
\end{equation} 
Equation $14$ is the intial condition and Equations $15,16$ are the boundary conditions at $x = 0$ and $x = L$ respectively. When $b_{11}= 0$ and $b_{12}= A(0)/R_{i}$, $f_{b1}(t)$ is a current injection at $x = 0$. When $b_{21} = 0$ and $ f_{b2}(t)= 0$, then it implies a sealed end boundary condition at $x = L$. \\
The initial conditions for equations $5,6,7$  take the form :\\
\begin{equation}
  y(0,V_{0}(x))= y_{\infty}(V_{0}(x))                                                           
\end{equation} 
All activation and inactivation variables start from the steady state belonging to the intial membrane potential distribution $V_{o}(x)$.\\
\section*{\begin{normalsize}\textbf{Configuration simulated}\end{normalsize}}
\begin{figure}[!ht]
\begin{center}
\subfigure[\textbf{Point soma dendrite construct with current injection at $i = 1$}]{
\includegraphics[width= 0.45\textwidth]{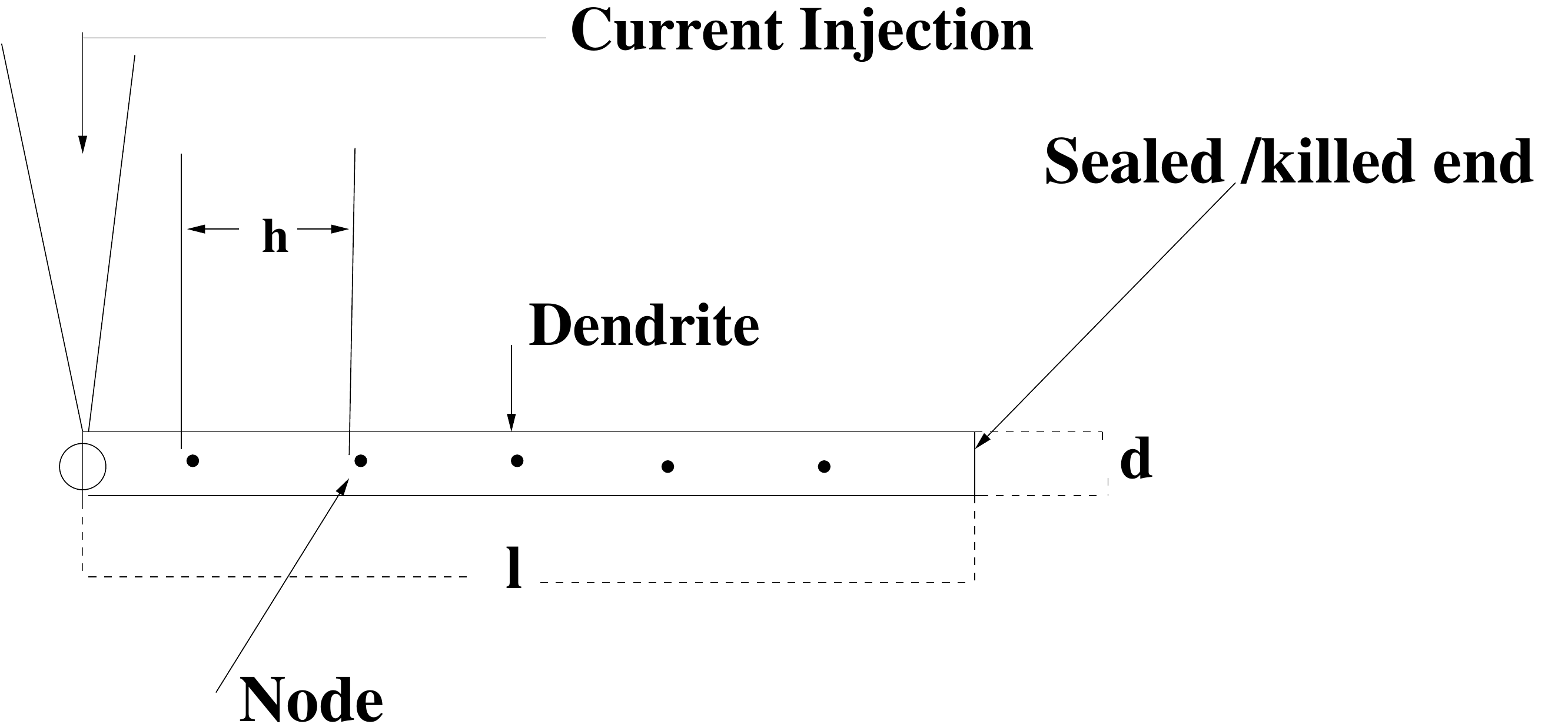}
}
\subfigure[\textbf{Equivalent circuit underlying the Hodgkin Huxley equations}]{
\includegraphics[width = 0.45\textwidth]{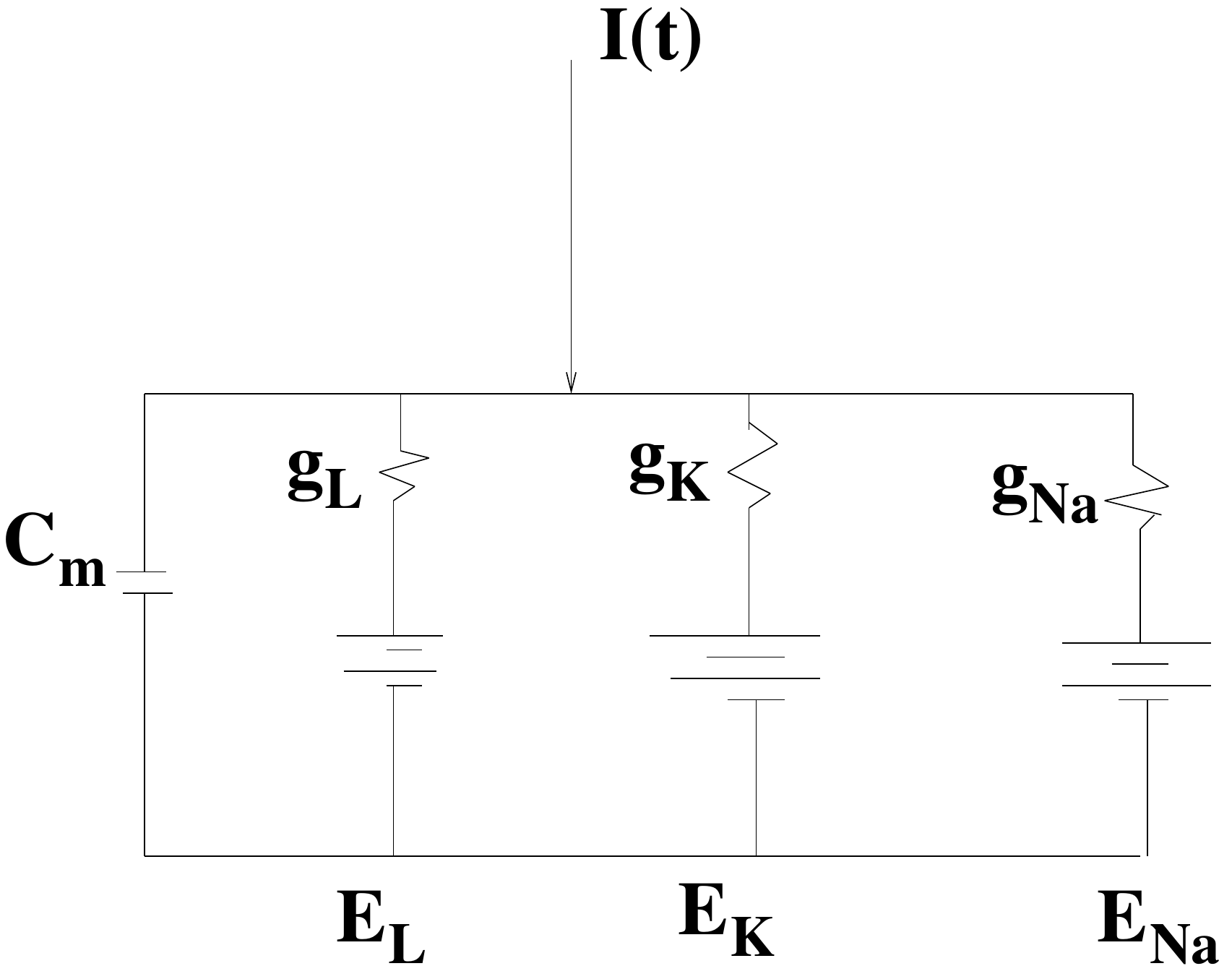}
}
\end{center}
\caption{\textbf{Soma dendrite construct}}
\label{fig:somadendact}
\end{figure}
\begin{figure}[!ht]
\subfigure[\textbf{Uniformly distributed channels}]{
\includegraphics[width = 0.45\textwidth]{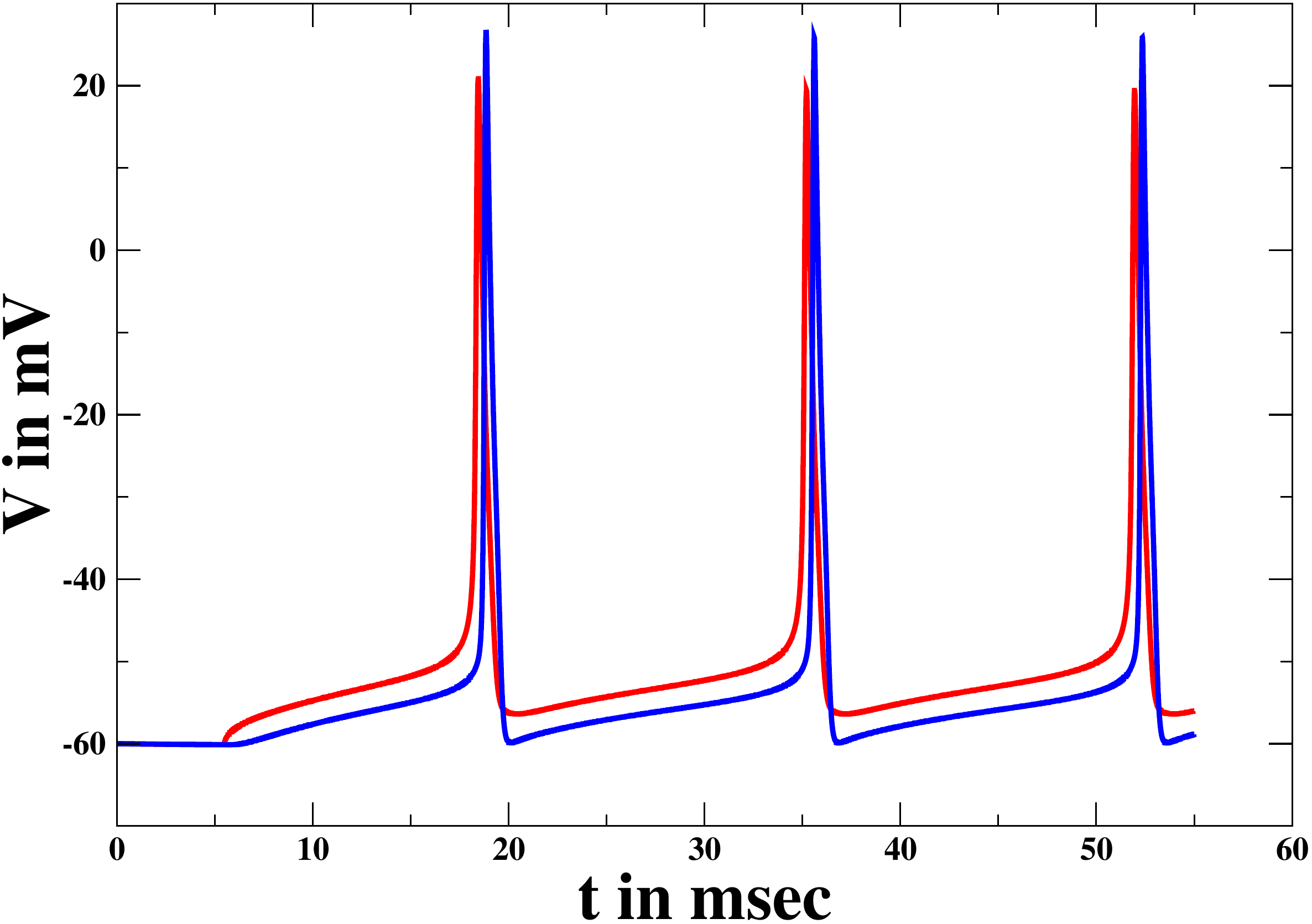}
}
\subfigure[\textbf{Exponentially distributed channels}]{
\includegraphics[width = 0.45\textwidth]{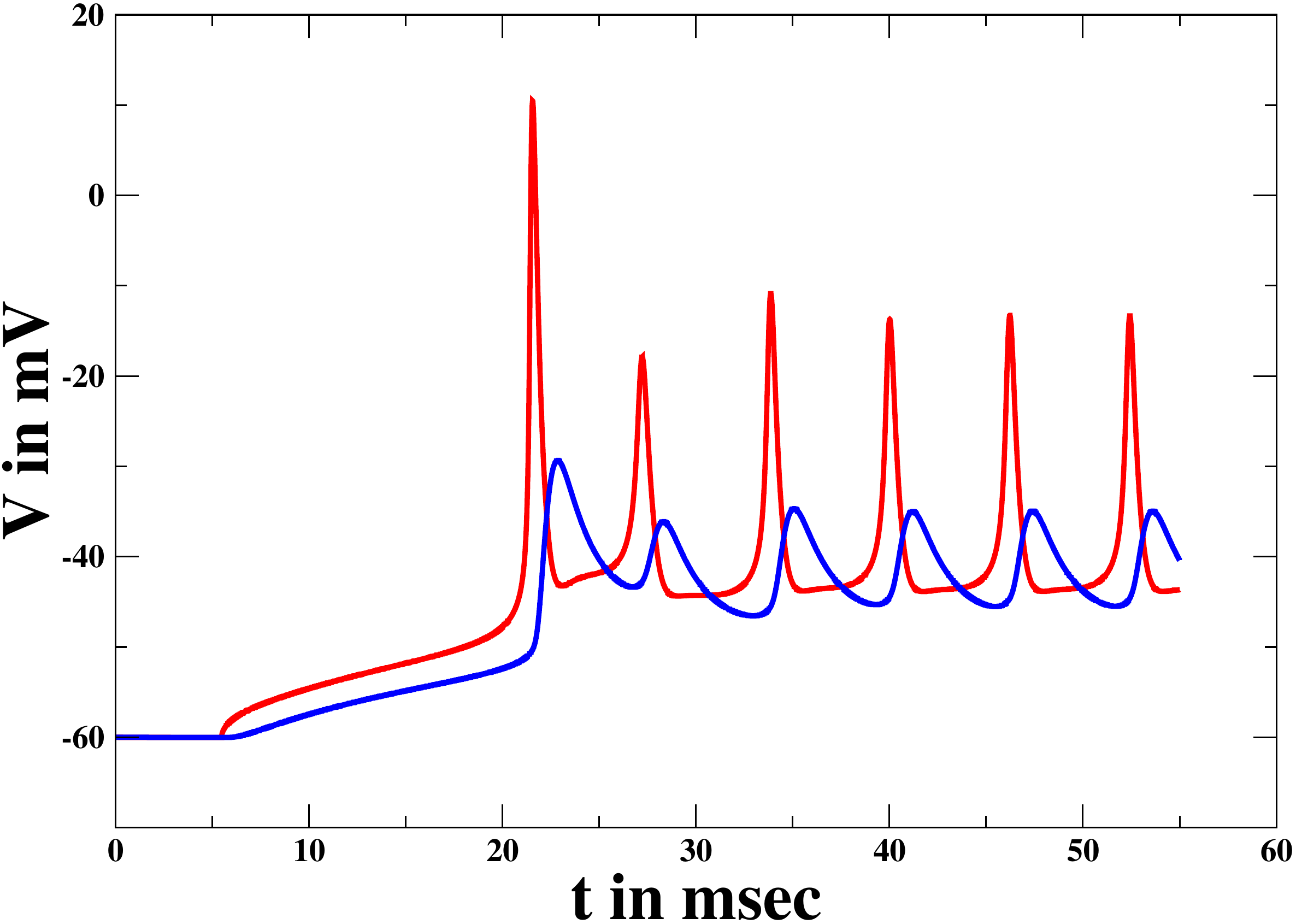}
}\\[3ex]
\subfigure[\textbf{Uniformly distributed K and exponentially distributed Na channels}]{
\includegraphics[width = 0.45\textwidth]{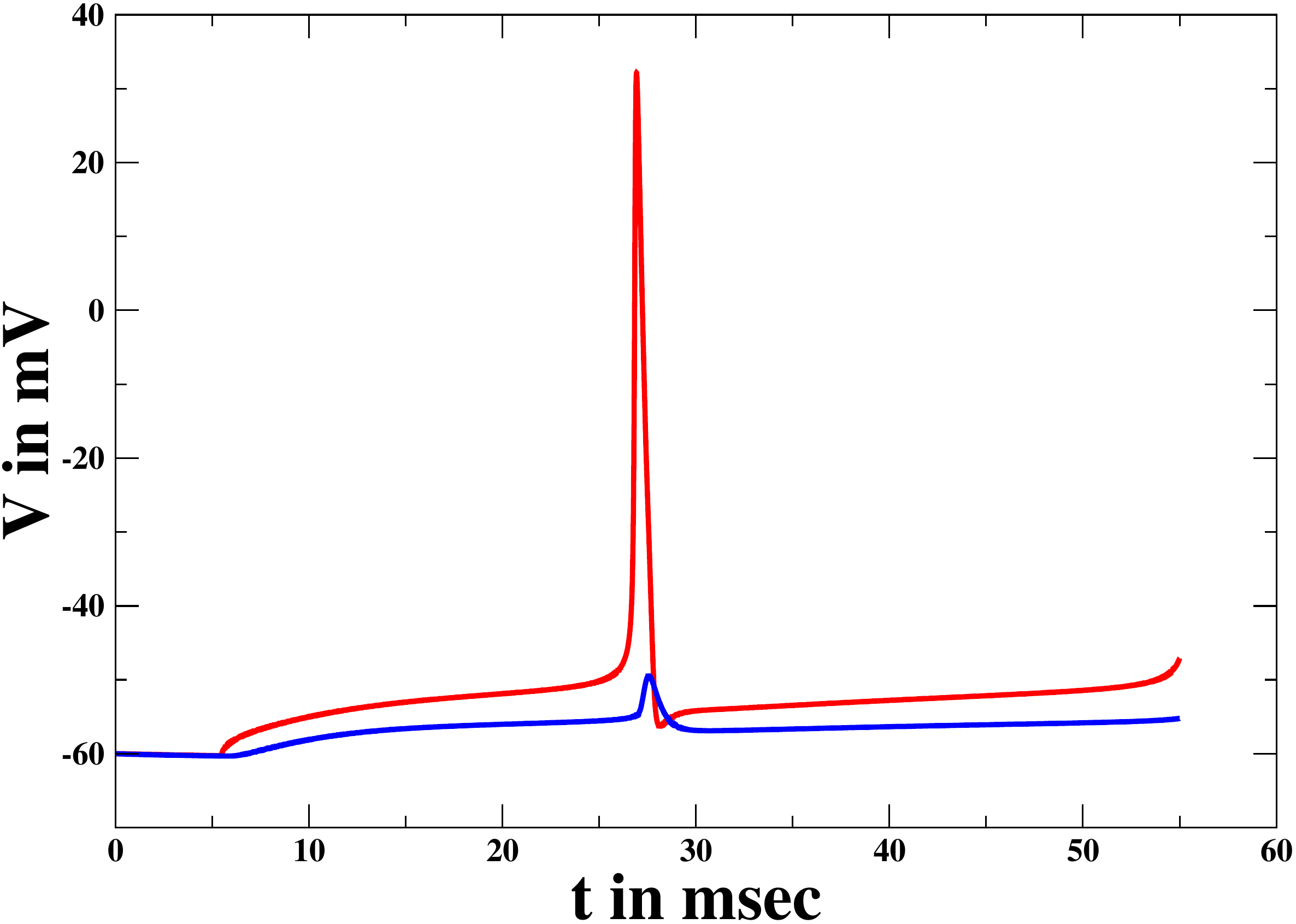}
}
\subfigure[\textbf{Exponentially distributed Na and K channels with change in leakage conductance in a small region}]{
\includegraphics[width = 0.45\textwidth]{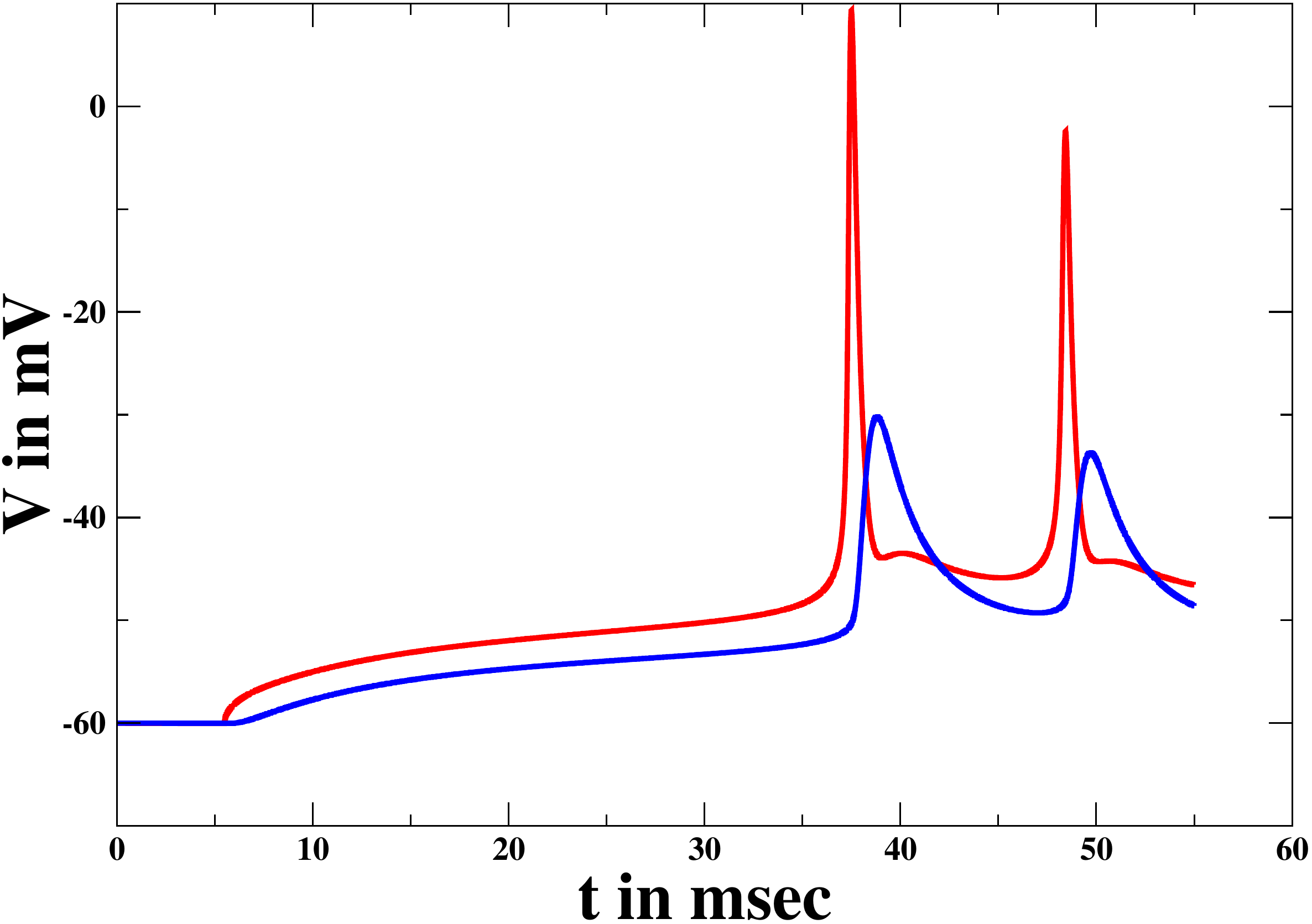} 
} 
\caption{\textbf{Simulation with four different dendritic distributions of voltage-gated ion channels.The same somatic stimulus ($450 \mu A/cm^{2}$)was applied to a cylindrical dendrite of radius $r = 1.85 \mu$m and length $ L= 400 \mu $m. Except for (d), the leakage channels were uniformly distributed. (a) uniformly distributed Na and K channels with $g_{Na} = g_{Na0} = 50mS/cm^{2}$, $g_{K} = g_{K0} = 12.5 mS/cm^{2}$.(b)exponentially distributed Na and K channels with $\lambda_{Na} = \lambda_{K} = -0.025 \mu$m$^{-1}$, and $g_{Na0}$, $g_{K0}$ are the same as in (a). (c) Uniformly distributed K channels as in (a), and exponentially distributed Na channels with the same $\lambda_{Na}$ as in (b) but with a largely increased maximal conductance, $g_{Na0} = 168 mS/cm^2 $ on the soma. (d) exponentially distributed Na and K channels with the same parameter values as in (b) but with the leakage conductance at $g_{L} = 0.5 mS/cm{^2}$ on the interval $[120,160]\mu$m. { soma, $ i = 1$,{\color{red} \textemdash} ;{end of dendrite,$ i = N$,{\color{blue} \textemdash }}}}}
\label{fig:cabunbact1}
\end{figure}
\begin{figure}[!ht]
 \subfigure[\textbf{Uniformly distributed channels}]{
\includegraphics[width = 0.45\textwidth]{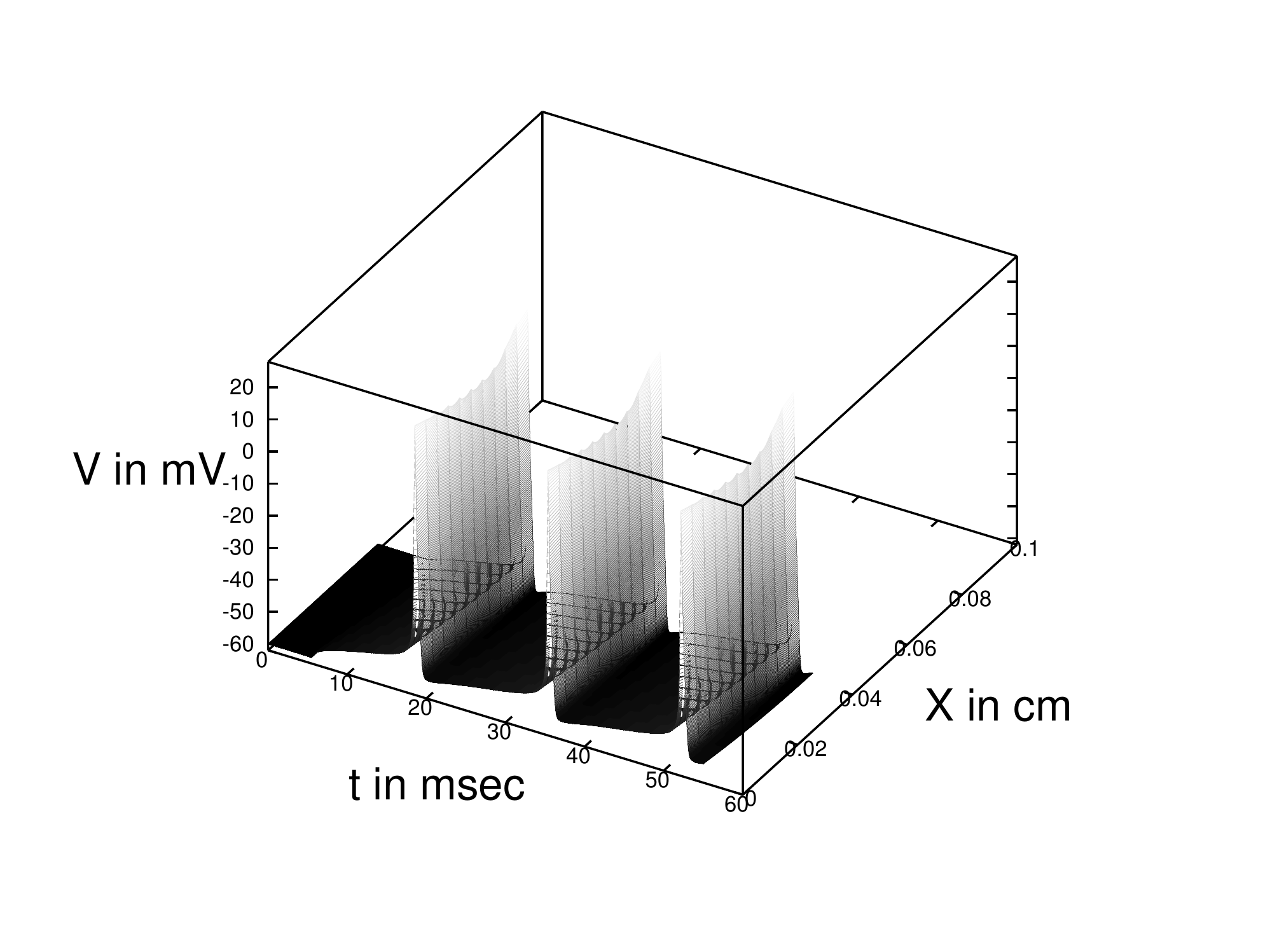}
}
\subfigure[\textbf{Exponentially distributed channels}]{
\includegraphics[width = 0.45\textwidth]{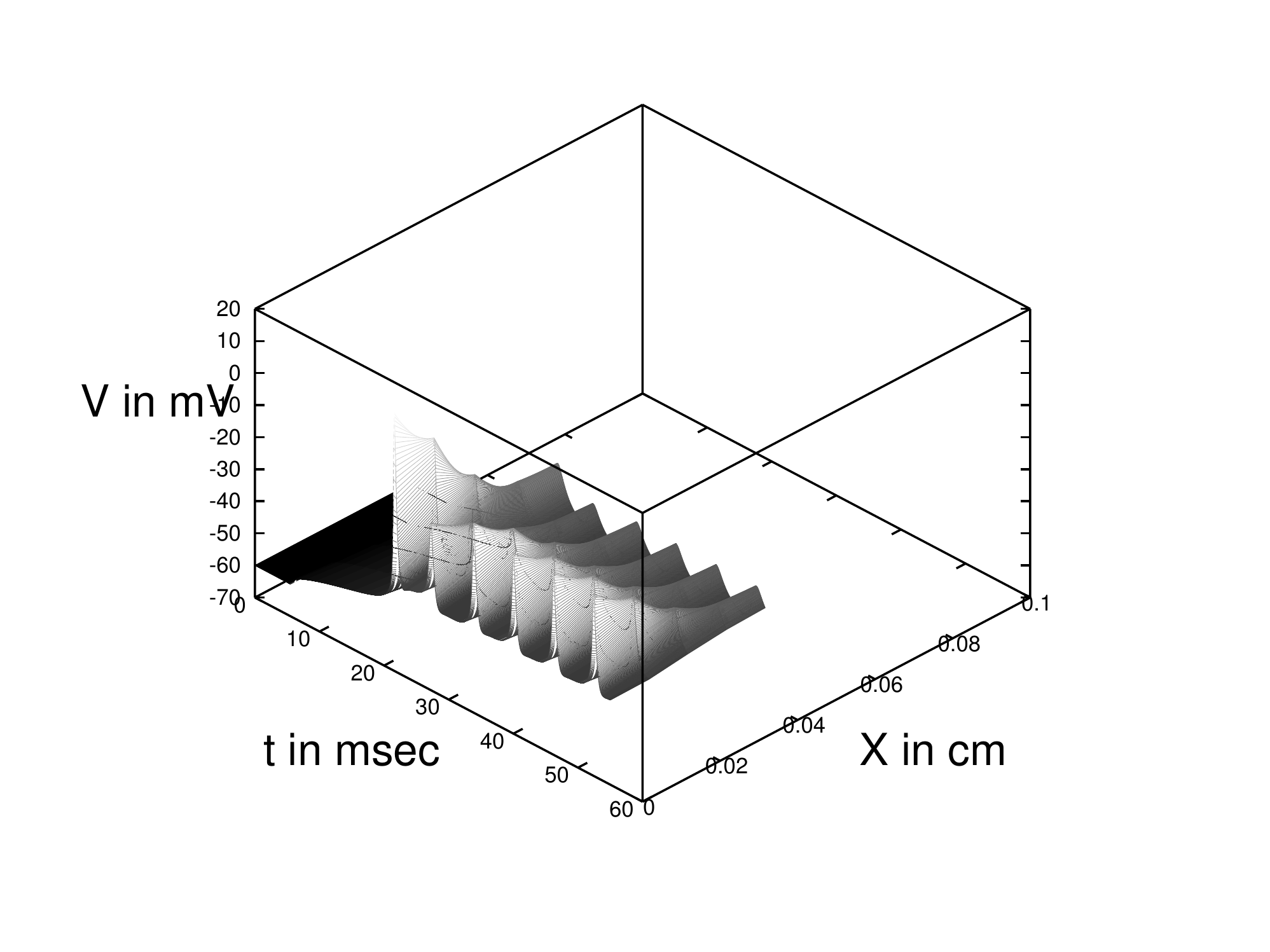}
}\\[3ex]
\subfigure[\textbf{Uniformly distributed K and exponentially distributed Na channels}]{
\includegraphics[width = 0.45\textwidth]{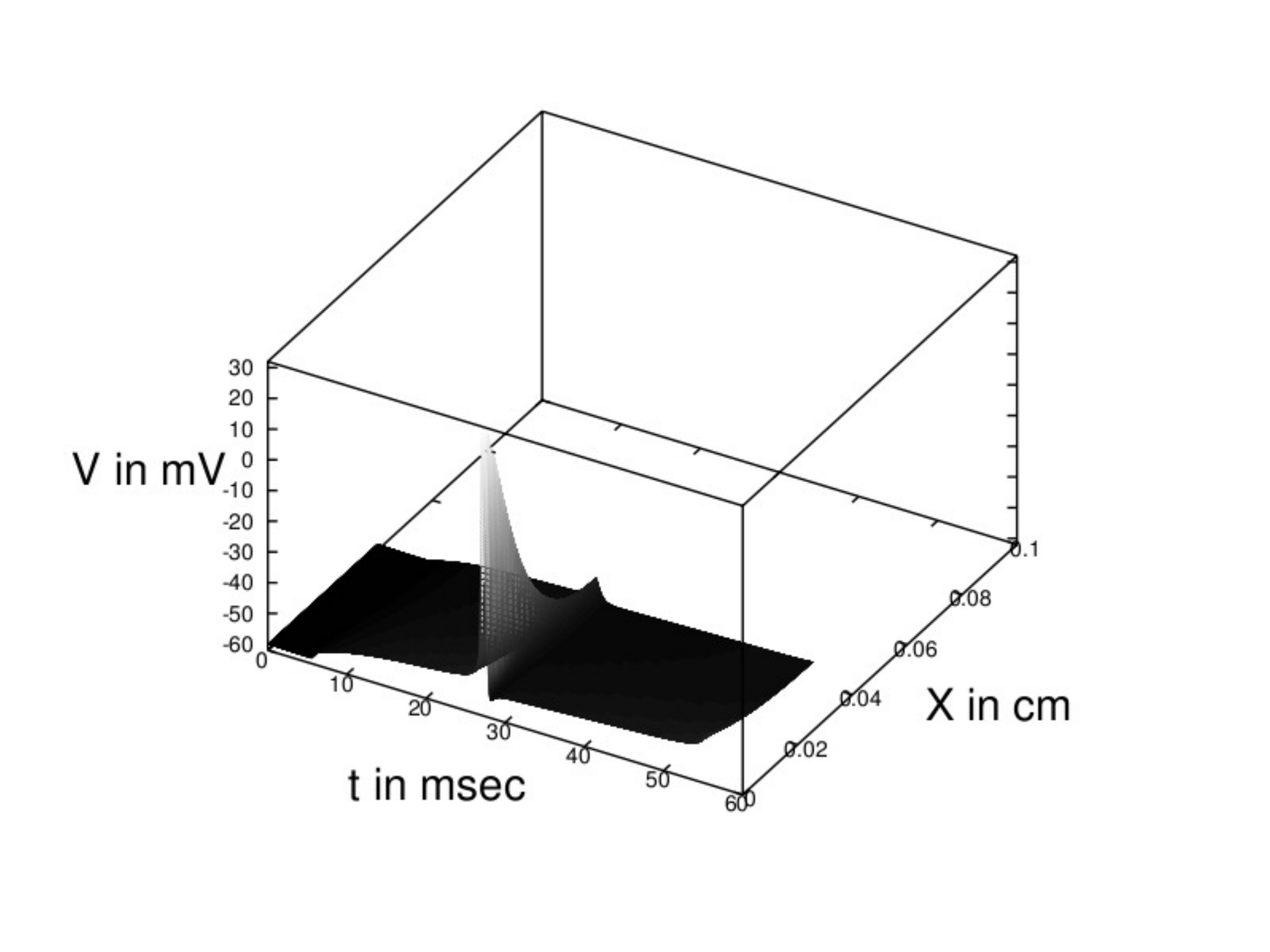}
}
\subfigure[\textbf{Exponentially distributed Na and K channels with change in leakage conductance in a small region}]{
\includegraphics[width = 0.45\textwidth]{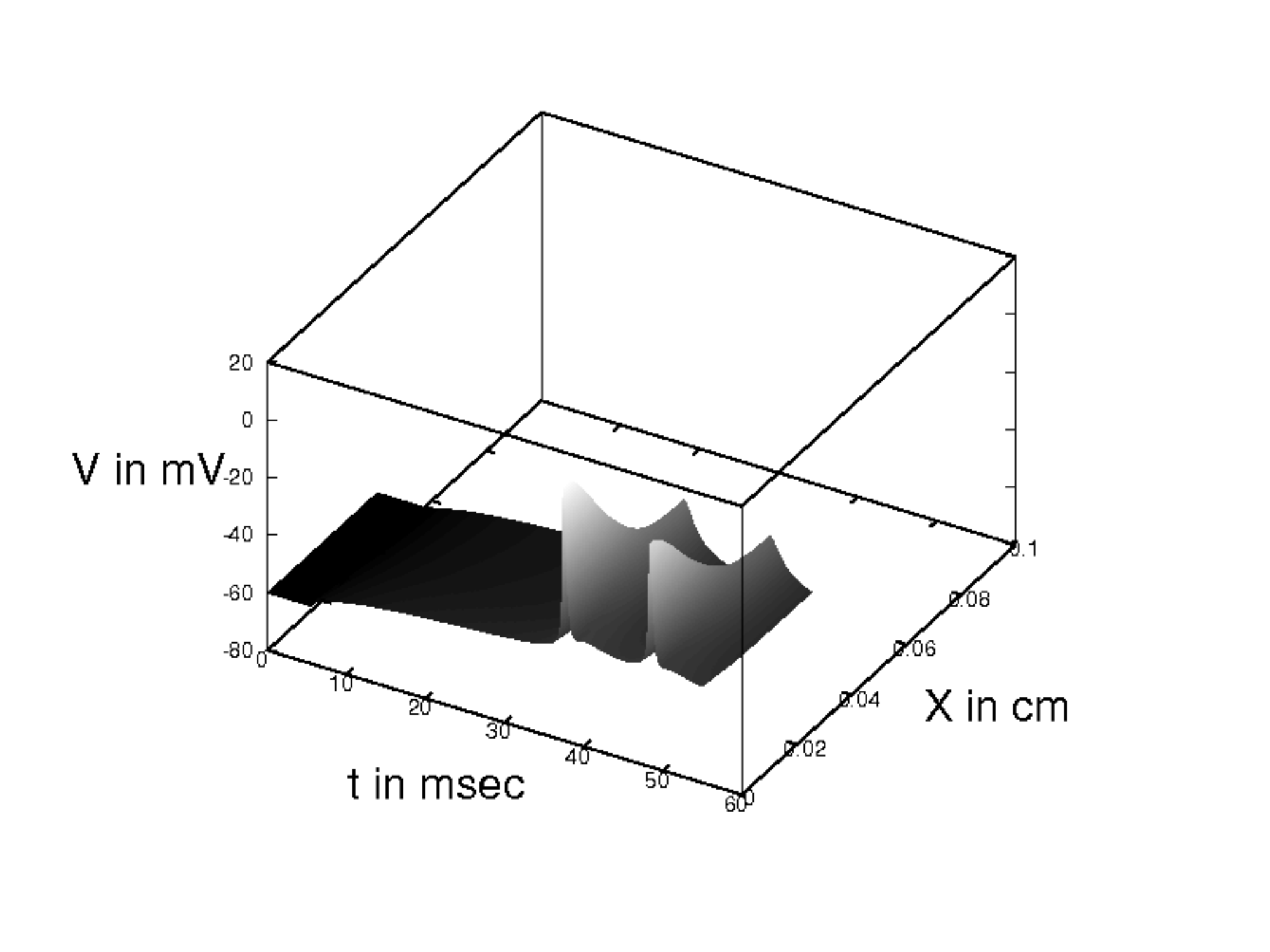} 
}
\caption{\textbf{Simulation with four different dendritic distributions of voltage - gated Na and K channels. Same protocol as in (Fig.~\ref{fig:cabunbact1}).}}
\label{fig:cabunbact1mesh}
\end{figure}

\parindent = 1in Simulation was done with four different dendritic distributions of voltage-gated ion channels. The current is injected at the point soma. (Fig.~\ref{fig:somadendact}) shows the model.(Fig.~\ref{fig:cabunbact1}) gives the results under the various conditions described in the legend. 
The four cases in (Fig.~\ref{fig:cabunbact1}) show that variations in distribution of Na and K channels lead to differences in firing in the dendrite. In unpublished results, it has been seen that changing the values of distribution of ion channels $\lambda _{Na} $ and $\lambda_{K}$ changes the firing properties. These are used to calculate $g_{Na}$ and $g_{K}$ in equation $4$. In (Fig.~\ref{fig:cabunbact1}) b and c, the value of $\lambda_{Na}$ and $\lambda_{K}$ used is $-0.015\times10^4$ cm$^{-1}$ unlike the $-0.025 \times 10^4 $ cm $^{-1}$ used in reference $14$. It can be seen that the results map the results produced in reference $14$ using spectral techniques. It can thus be argued that the compact difference scheme is useful in solving the HH formulations.
(Fig.~\ref{fig:cabunbact1mesh})gives the 3-D plot for the various conditions described in (Fig.~\ref{fig:cabunbact1}). 
\clearpage
\section*{Tables}
\begin{table}[h!]
\caption{\bf{Values of $\Delta T$ at various $N$}}
\centering
\begin{tabular}{c rr}
\hline 
$N$& \multicolumn{2}{c}{$\Delta T$} \\
\hline
$11$ & $1441.4$\\
$21$& $360.36$ \\
$41$ & $90.089$  \\
$61$ & $40.040$ \\
$81$ & $22.522$\\
\hline
\end{tabular} 
\label{tab:DeltaT}
\end{table}
\begin{table}[h!]
\caption{\bf{Parameters of dendrite used in simulation}}
\centering
\begin{tabular}{c rr}
\hline 
Parameter& \multicolumn{2}{c}{Values} \\
\hline
length & $400 \mu$\\
diameter& $3.7 \mu$ \\
$R_{m}$ & $10^{-3 }\Omega$.cm$^{2} $  \\
$R_{i}$ & $330  \Omega$.cm \\
$C_{m}$ & $1^-6 Farad/cm^{2} $\\
$\tau$ & $1^{-5}$ sec \\
$Iinj$ & $4.8384^{-11} $amperes \\
$V_{Na}$& $55 $ mV \\
$V_{K}$ & $ -95$ mV \\
$V_{L}$&$-60$ mV\\
\hline
\end{tabular} 
\label{tab:parameters}
\end{table}
\begin{table}[h!]
\caption{Resolving Efficiency $\epsilon$ of the second derivative schemes,($15$, Table $5$)}
\centering
\begin{tabular}{c c c c }
\hline\hline
Scheme & $\epsilon = 0.1$ & $\epsilon = 0.01$ & $\epsilon = 0.001$  \\[0.5ex]
\hline
Fourth order central & 0.59& 0.31 & 0.17\\
Fourth order compact  & 0.68 & 0.39 & 0.22 \\
Sixth order tridiagonal & 0.80 & 0.55 & 0.38 \\
\hline 
\end{tabular}
\label{tab:resolving}
\end{table}
\section*{\begin {normalsize}\textbf{Convergence}\end{normalsize}} The solution converges to a constant value as N is increased. The peak to peak distance recorded at $x=0$ is used to measure convergence. As fourth order compact and fourth order central difference schemes are compared, it is seen that the convergence rate in $1a$ is almost the same for both. In 
$1b$, $1c$ and $1d$ the solution from the compact scheme converges at a slightly faster rate than the central difference scheme. Decreasing $t$ to $t/2$ and $t/4$ does not change the solution. 
This is described in (Table~\ref{tab:conver}) and Figures in Supplementary Information. 
\section*{\begin{normalsize}\textbf{Concluding Remarks}\end{normalsize}} In this paper the compact difference scheme is used to solve the Hodgkin-Huxley equations. This is an alternative to the solution using Chebyshev approximations $^{14}$. The authors had discussed the advantages of the Chebyshev method - high numerical accuracy at any spatial point, negligible truncation errors leading to errors only from the rounding errors;space dependent parameter distributions as well as mixed linear boundary conditions could be implemented. In this paper using compact difference scheme, we have shown that the same can be achieved with far greater ease. As shown above, the solutions to changes in ion channel densities can be obtained. However implicit schemes for time stepping are difficult to implement using this scheme. This disadvantage can be overcome with using the corrector- predictor method.Table~\ref{tab:resolving} gives the resolving efficiency of the second derivative schemes at three different errors, $\epsilon$. (Reference $15$ defines resolving efficiency as the modified wavenumbers of the spectral scheme and the differencing scheme.) This is specific to a scheme for any given error. It can be seen at $ \epsilon <= 0.001$ , the resolving efficiency of the fourth order compact scheme is $0.22$ which is greater than that of the fourth order central scheme which is $ 0.17$. It is also seen that the resolving efficiency of the sixth order compact scheme is $0.38$ while that of explicit sixth order central differences is $0.29$. Furthermore an eighth order tridiagonal scheme can yield at $ \epsilon <= 0.001$, a resolving efficiency of $0.48$ and a spectral like pentadiagonal scheme can yield a resolving efficiency of $0.84$. In a paper (ref.$ 22 $), Gopinathan has compared solutions of the passive cable equation solved with the compact scheme  with analytical equations for both sealed and killed end and found that even as low as $ N=10$, there is complete overlap. In the absence of analytical equations for the HH formulation, comparisons have been made with the spectral methods based solution in reference ${14}$ while convergence has been tested using comparisons between fourth order central  and compact schemes. The changed values of $\lambda _{Na}$ and $\lambda_{K}$ discussed earlier apply to both compact and central schemes to yield results similar to that in reference $14$.Thus it does not seem scheme specific.  Ongoing work looks at using the compact scheme to solve HH equations for tapered and branched dendrites. While, the compact difference scheme is used increasingly to advantage in calculations of turbulence in airflow, to the best of our knowledge this is the first time it has been applied to study changes in electrical flow in dendrites  in the brain. This is a powerful tool which can be modified and applied to a range of equations where spatial resolution is significant.    
\section*{\begin{normalsize}\textbf{References}\end{normalsize}}
1.Fiale,J.C,Spacek,J \& Harris,K.M in \textit Dendrites( ed.Stuart,G, Spruston,N \& Hausser,M)1-41,(Oxford University Press, 2008)\\
2. Yuste,R in \textit Dendritic Spines,(MIT Press,2010)\\
3. Segev,I \& Rall,W. Excitable dendrites and spines:earlier theoretical insights elucidate recent direct observations,\textit Trends in Neuroscience,\textbf{21}, 453-460,(1998)\\
4. Yuste,R. Dendritic Spines and Distributed Circuits,\textit Neuron,\textbf{71},772-782,(2011)\\
5. Stuart,G \& Hausser,M.Initiation and spread of sodium action potentials in cerebellar Purkinje cells,\textit Neuron,\textbf{13},703-712,(1994)\\
6. Stuart,G \& Sakmann,B. Active propagation of somatic action potentials into neocortical pyramidal cell dendrites,\textit Nature,\textbf{367},69-72,(1994)\\
7. Stuart,G,Schiller,J \& Sakmann,B.Action potential initiation and propagation in rat neocortical pyramidal neurons,\textit J Physiol(Lond),\textbf{505},617-632,(1997a)\\
8. Stuart,G ,Spruston,N,Sakmann,B \& Hausser,M.Action potential initiation and backpropagation in neurons of the mammalian CNS,\textit Trends in Neuroscience,\textbf{20},125-131,(1997b)\\
9. Magee,J.C in \textit Dendrites ( ed.Stuart,G ,Spruston,N \& Hausser,M)225-250,(Oxford University Press, 2008)\\
10. Rall,W. Branching dendritic trees and motoneuron membrane resisitivity,\textit Exptl Neurol,\textbf{1},491-527,(1959)\\
11.Koch,C in \textit Biophysics of Computation -information processing in single neurons, chapters 2,6,(Oxford University Press,1999)\\
12.Tuckwell,H.C in \textit Introduction to Theoretical Neurobiology-Linear Cable Theory and Dendritic Structure ,124-179,(Cambridge University Press,1988)\\
13.Hodgkin,A.L and Huxley,A.F. A quantitative description of membrane current and its application to conduction and excitation in nerve.\textit J Physiol\textbf{1},491-527,(1952c)\\
14.Toth,T.I and Crunelli,V. Solution of the nerve cable equation using Chebyshev approximations.\textit Journal of Neuroscience Methods,\textbf{87},119-136,(1999)\\
15.Lele,S.K.Compact finite difference schemes with spectral-like resolution.\textit J.Comp Phy,\textbf{103},16-42,(1992)\\
16. Moin,P in \textit Fundamentals of Engineering Numerical Analysis ( Cambridge University Press,2001)\\
17.Lindsay,K, Ogden,J, Halliday,D.M.\& Rosenberg,J.R in \textit Modern Techniques in Neurosciences Research (ed.Windhorst U \& Johansson H),213-306,(Springer-Verlag,1999)\\
18. Lindsay,K.A. and Ogden,J.M. and Rosenberg,J.R. in \textit Biophysical Neural Networks ( ed. Poznanski,R.R),Chapter 15,( Mary Ann Liebert Inc., 2001)\\
19. Toth T I and Crunelli V. Effects of tapering geometry and inhomogeneous ion channel distribution in a neuron model.\textit Neuroscience,\textbf{84:4},1223-1232,1998 \\
20. Bower J M and Beeman D in \textit The Book of GENESIS : Exploring Realistic Neural Models with the GEneral NEural SImulation System ( Springer- Verlag,1998) \\
21. Carnevale N T and Hines M L in \textit The NEURON book ( Cambridge University Press, 2006)\\
22.Gopinathan,A. Solving the cable equation using a compact difference scheme - passive soma dendrite.(http://arxiv.org/abs/1301.2885)\\
\textbf{Supplementary Information} is available in the online version of the paper.\\
\textbf{Acknowledgements} AG would like to acknowledge the support provided by A.K Gupta( currently NIMHANS, Bangalore)and M.D Nair of Sree Chitra Tirunal Institute of Medical Sciences and Technology, Tiruvananthapuram.  AG would also like to acknowledge the support of  V.Nanjundiah in making arrangements to work at the Indian Institute of Science. AG thanks Elizabeth Jacob for support provided at NIIST, Trivandrum during the writing of this paper. AG thanks Maya Ramachandran and Venugopalan for acquiring necessary references from the library of the National University of Singapore. Thanks are due to Ganesh of SPACE, Tiruvananthapuram for help with opensource software.AG also acknowledges the ready help provided by the octave- users group in solving any problems that have arisen with the Octave code. AG is supported by the DST- WOS-A grant which covered the costs of this project. \\
\textbf{Author Contributions} JM suggested the use of the compact difference scheme as an alternative to the spectral scheme and helped in smoothing out troubles during its implementation. AG ran the simulations, wrote the code and wrote the paper. Both authors interpreted the results and edited the papers. \\
\textbf{Author Information} Reprints and permission information is available at www.nature.com/reprints. The authors declare no competing financial interests. Readers are welcome to comment on the online version of the paper. Correspondence and requests for materials should be addressed to AG(dendron.15@gmail.com)\\
\textbf{Supplementary Information}
\section*{\begin{normalsize}\textbf{Supplementary Methods}\end{normalsize}}
The following figure (Fig.~\ref{fig:cabunbactcent1}) illustrates the difference in convergence of compact and central differences ( fourth order for four different cases). The smallest value of N is chosen as the least one where the cell fires. 
\begin{figure}[!ht]
\subfigure[\textbf{Uniformly distributed channels}]{
\includegraphics[width = 0.45\textwidth]{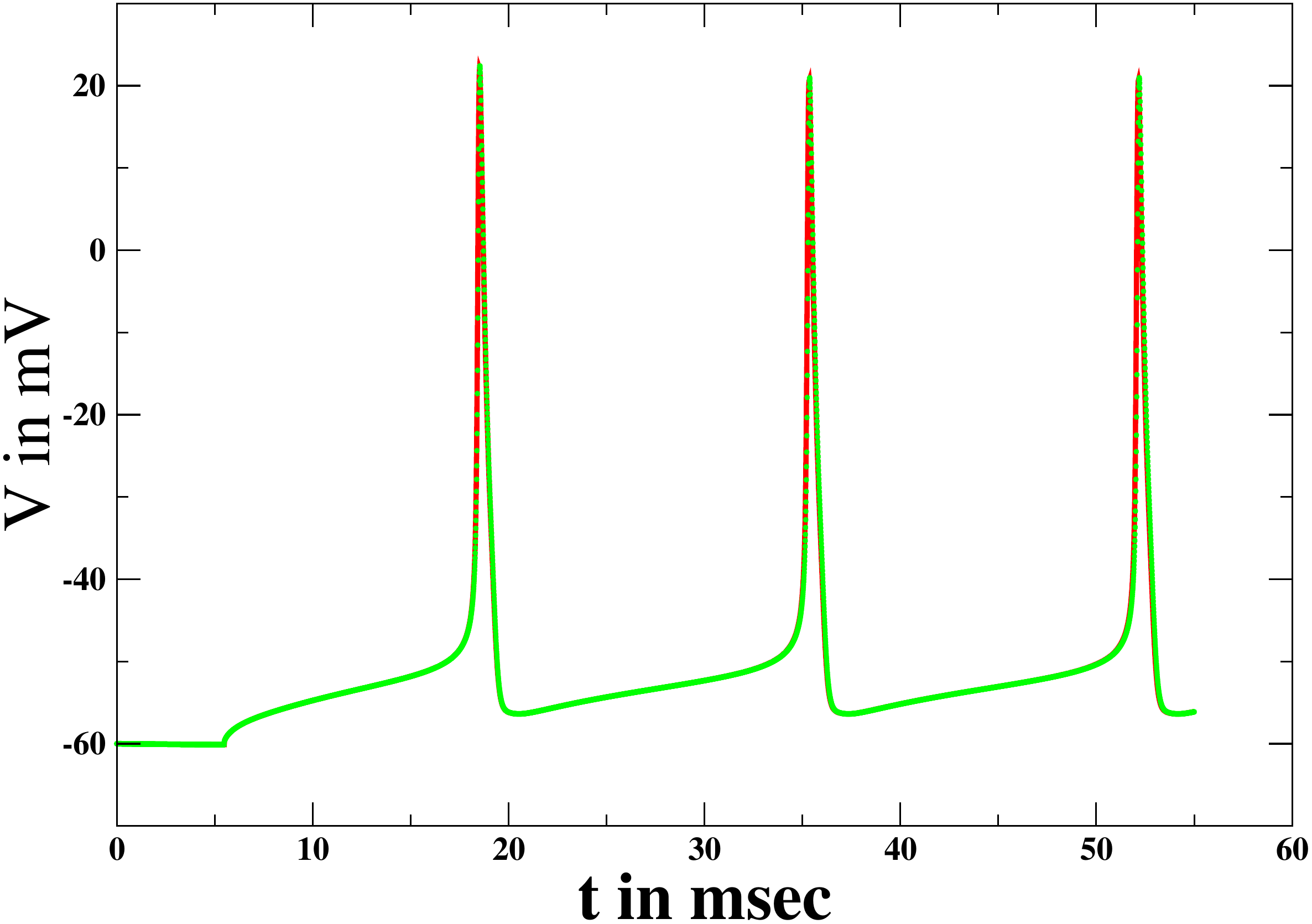}
}
\subfigure[\textbf{Exponentially distributed channels}]{
\includegraphics[width = 0.45\textwidth]{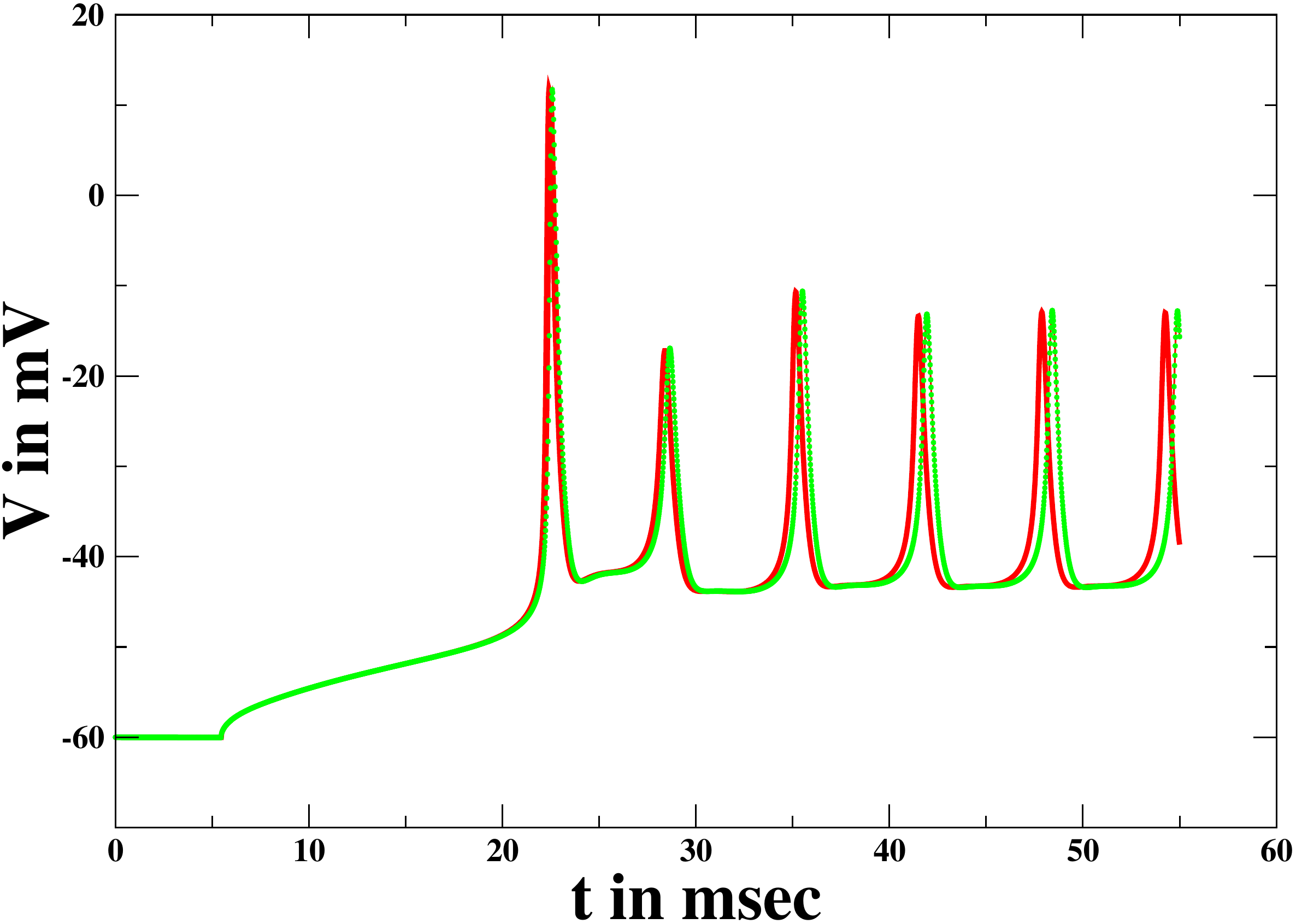}
}\\[3ex]
\subfigure[\textbf{Uniformly distributed K and exponentially distributed Na channels}]{
\includegraphics[width = 0.45\textwidth]{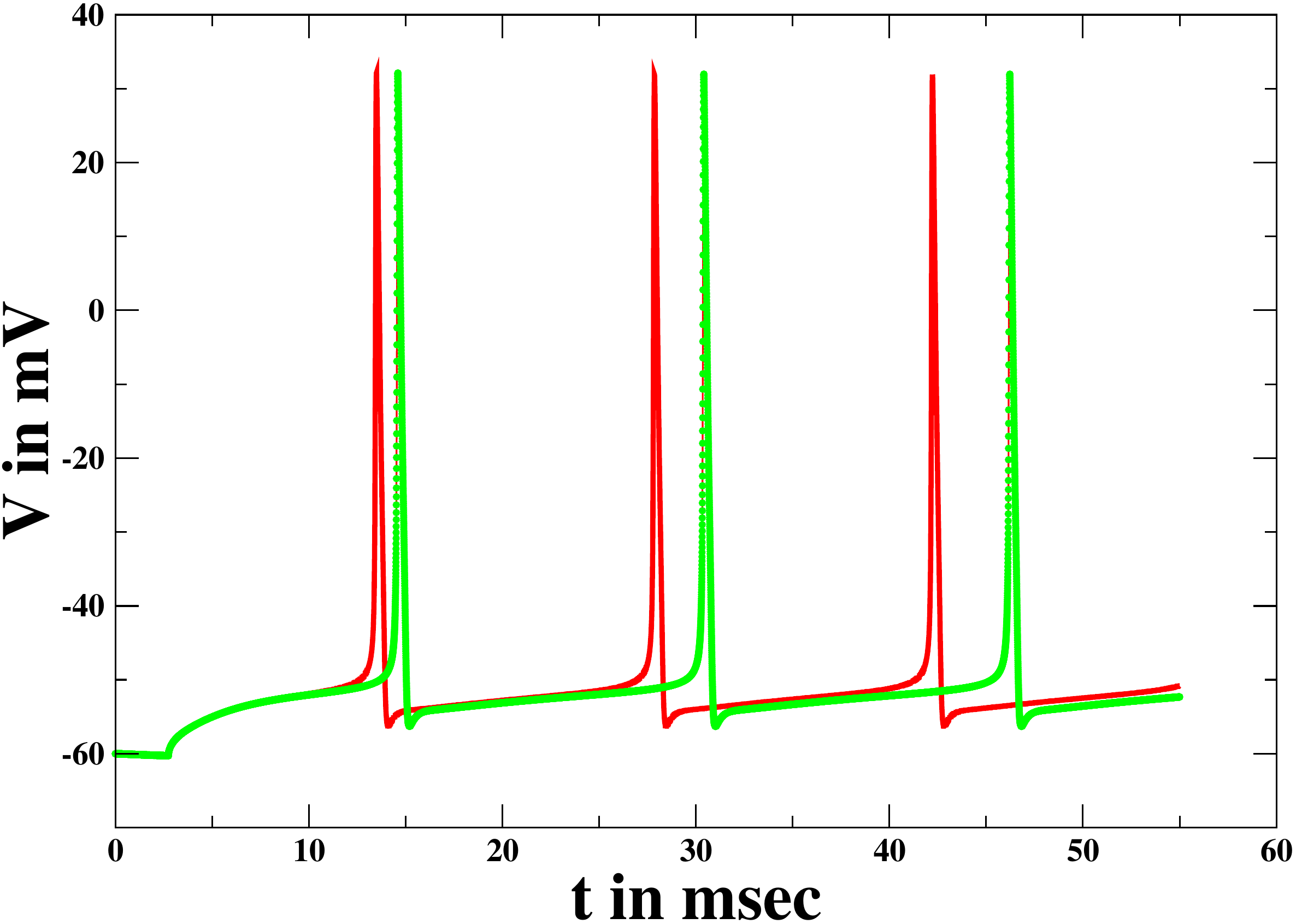}
}
\subfigure[\textbf{Exponentially distributed Na and K channels with change in leakage conductance in a small region}]{
\includegraphics[width = 0.45\textwidth]{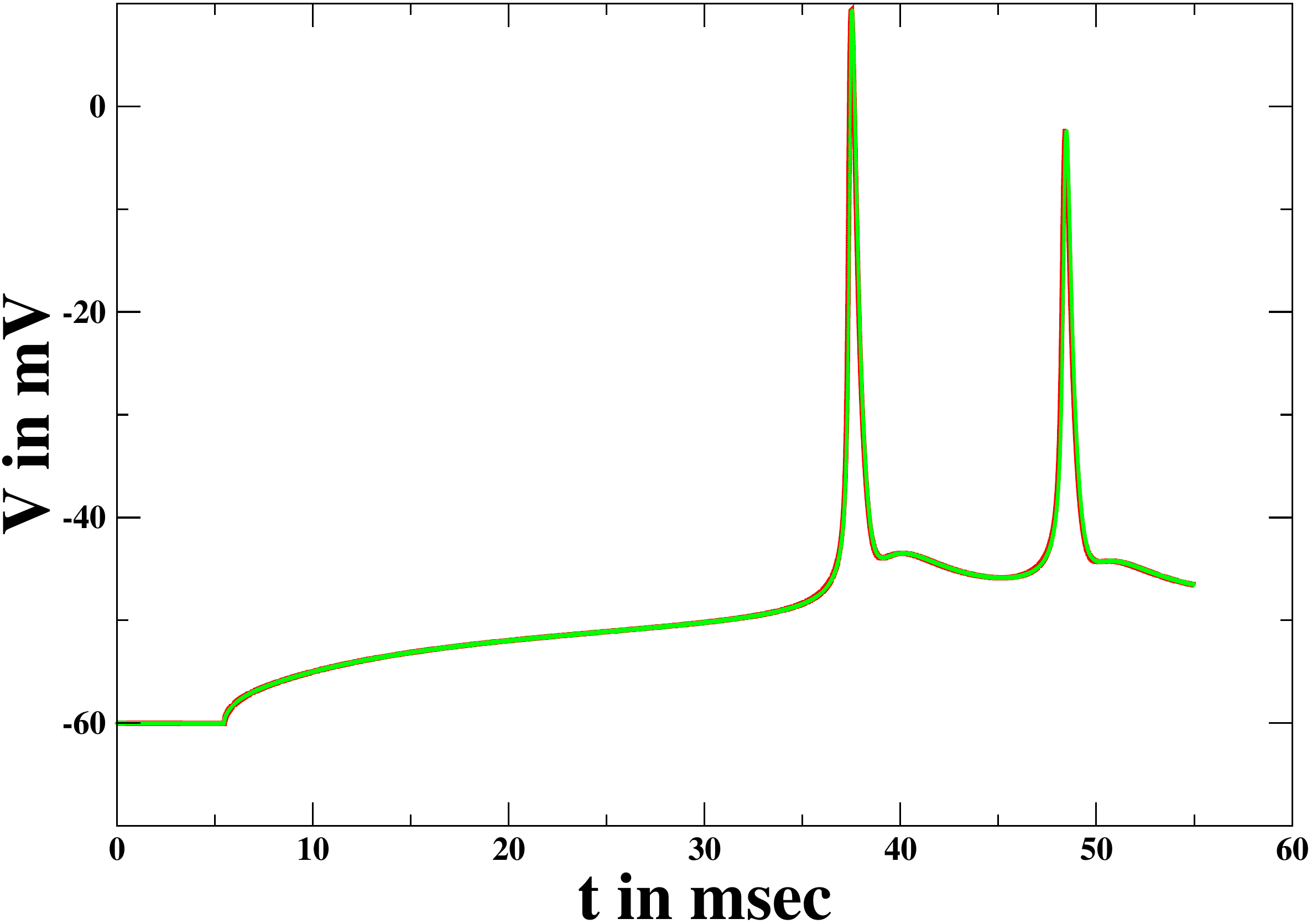} 
} 
\caption{\textbf{Simulation with four different dendritic distributions of voltage-gated ion channels.The same somatic stimulus ($450 \mu A/cm^{2}$)was applied to a cylindrical dendrite of radius $r = 1.85 \mu$m and length $ L= 400 \mu$ m. Except for (d), the leakage channels were uniformly distributed. (a) uniformly distributed Na and K channels with $g_{Na} = g_{Na0} = 50mS/cm^{2}$, $g_{K} = g_{K0} = 12.5 mS/cm^{2}$.(b)exponentially distributed Na and K channels with $\lambda_{Na} = \lambda_{K} = -0.025 \mu$m$^{-1}$, and $g_{Na0}$, $g_{K0}$ are the same as in (a). (c): Uniformly distributed K channels as in (a), and exponentially distributed Na channels with the same $\lambda_{Na}$ as in (b) but with a largely increased maximal conductance, $g_{Na0} = 168 mS/cm^{2}$ on the soma. (d) exponentially distributed Na and K channels with the same parameter values as in (b) but with the leakage conductance at $g_{L} = 0.5 mS/cm^{2}$ on the interval $[120,160]\mu$m. { compact,{\color{red} \textemdash} ;{central,{\color{green} \textemdash }}}}}
\label{fig:cabunbactcent1}
\end{figure} 
\oddsidemargin -2.0cm
\evensidemargin -2.0cm
\begin{table}[h!]
\caption{Convergence of interpeak distance for (Fig.~\ref{fig:cabunbact1})}
\centering
\begin{tabular}{c c c c}
\hline\hline
Problem & $N$ & $compact$ & $central$ \\[0.5ex]
\hline
1a & 11& 16.808 & 16.824 \\ 
   & 21 & 16.762 & 16.764 \\
   & 41 & 16.748 & 16.748 \\
1b & 11 & 7.4369 & 7.5638 \\
   & 21 & 6.9134 & 6.9451\\
   & 41 & 6.7854 & 6.7903 \\
1c & 21 & 14.366 & 15.811 \\
   & 41 & 10.215 & 10.307 \\
1d & 41 & 10.930 & 10.946\\
   & 81 & 9.1056 & 9.1063\\
\hline 
\end{tabular}
\label{tab:conver}
\end{table}
\end{document}